\documentclass{emulateapj}
\usepackage{natbib,graphicx}

\shorttitle{Early-type host galaxies of Type II/Ib Supernovae}
\shortauthors{Suh et al.}

\begin{document}
\title{Early-type Host galaxies of Type II and Ib Supernovae}

\author{Hyewon Suh\altaffilmark{1}, Sung-chul Yoon\altaffilmark{2}, Hyunjin Jeong\altaffilmark{1}, Sukyoung K. Yi\altaffilmark{1}}
\altaffiltext{1}{Department of Astronomy, Yonsei University, Seoul 120-749, Korea; yi@yonsei.ac.kr} 
\altaffiltext{2}{Argelander Institute for Astronomy, University of Bonn, Auf dem Huegel 71, D-53121 Bonn, Germany} 

\begin{abstract}
Recent studies find that some early-type galaxies host Type II or Ibc supernovae (SNe II, Ibc).
This may imply recent star-formation activities in these SNe host galaxies,
but a massive star origin of the SNe Ib so far observed in early-type galaxies
has been questioned because of their intrinsic faintness and unusually strong Ca lines
shown in the nebular phase. To address the issue,
we investigate the properties of early-type SNe host galaxies
using the data with {\it Galaxy Evolution Explore}({\it GALEX}) ultraviolet
photometry, and the Sloan Digital Sky Survey (SDSS) optical data.
Our sample includes eight SNe II and one peculiar SN Ib (SN 2000ds) host galaxies
as well as 32 SN Ia host galaxies. The host galaxy of SN 2005cz, another
peculiar SN Ib, is also analysed using the {\it GALEX} data and the 
NASA/IPAC Extragalactic Database (NED) optical data.
We find that the NUV$-$optical colors of SN II/Ib host galaxies are
systematically bluer than those of SN Ia host galaxies, and some SN II/Ib
host galaxies with NUV$-$r colors markedly bluer than the others exhibit strong
radio emission.  We perform a stellar population synthesis analysis and find a
clear signature of recent star-formation activities in most of the SN II/Ib
host galaxies.  Our results generally support the association of the SNe II/Ib
hosted in early-type galaxies with core-collapse of massive stars.  We briefly
discuss implications for the progenitors of the peculiar SNe Ib 2000ds and
2005cz.  \\ \end{abstract}

\keywords{galaxies: elliptical and lenticular, cD -- galaxies: evolution -- supernovae: general}

\section{Introduction}
\label{sec:intro}

There is a general consensus that most (if not all) SNe II, Ib and Ic originate from
the core-collapse of massive stars. Observations show that they occur predominantly
in late-type galaxies where active star-formation takes place
\citep[e.g.,][]{Mvan76, Cap99, Hamuy03}.
However, several studies report observations of SNe II and Ibc in
early-type galaxies \citep{van02, van03, van05, Hakobyan08, Leaman10}, which
are traditionally considered old stellar populations
composed mainly of low-mass stars. Although \citet{Hakobyan08} argue that many of
the previously reported early-type SN II/Ib host galaxies may be misclassified
spirals, it is clear that at least some of them belong to genuine
early-types in the morphological sense \citep{Hakobyan08, Leaman10}. 

\citet{Perets10} note that all of the SNe Ibc observed so far in early-type
galaxies belong to a subclass of Type Ib supernovae; they show a typical
Type Ib like spectrum characterized by the strong helium lines and weak silicon
lines at early time, but they appear to be systematically fainter than
typical SNe Ib, and show strong Ca II and weak O I
emission lines in late-time optical spectra. Their origin is currently a
matter of debate.  Their progenitors might be related to rather old stellar
populations (e.g., helium-accreting white dwarfs or mergers of ONeMg- and He
white dwarfs; \citealt{Kawabata10, Perets10}), or their relatively frequent
detection in early-type galaxies might be due to their intrinsic faintness,
even if they have a core-collapse origin \citep{Kawabata10}. 

Given this intriguing finding, a careful study of the stellar population in
the early-type host galaxies of SNe II/Ibc is important for a better
understanding of recent star formation activities that may be related to such
events. The {\it Galaxy Evolution Explorer} ({\it GALEX}) ultraviolet (UV)
filters are particularly useful for this purpose, since they allow us to detect
even a tiny amount of young massive stars in galaxies. Many
recent studies with \emph{GALEX} data show that a significant fraction of
early-type galaxies exhibit enhanced UV light as a sign of recent star
formation \citep{Yi05, Salim07, Donas07, S07a, Kaviraj07, Kaviraj08}.  

In this paper, we present the UV$-$optical color-magnitude relation of
early-type host galaxies of some SNe II/Ib, including SN 2000ds and SN 2005cz,
which belong to the faint, Ca-rich class.  By comparing them to
early-type host galaxies of SNe Ia, we discuss whether the properties of
early-type SN II/Ib host galaxies differ systematically from those of SNe Ia,
and if the faint, Ca-rich class of SN Ib can still be explained within the framework of
the core-collapse scenario.

Throughout this paper we assume a $\Lambda$CDM cosmology with
$\Omega_{m} = 0.3$ and H$_{0} = 70$ km s$^{-1}$ Mpc$^{-1}$.


\begin{center}
\begin{deluxetable*}{clcrcccclcrcc}
\tabletypesize{\scriptsize}
\tablecaption{Sample \label{tbl:sample}}
\tablewidth{0pt}
\tablehead{
& \colhead{Host galaxy} &  &  &  & & & \colhead{Supernova} &  &  &\\
& & & & & & & & & & \\
\colhead{List} & \colhead{Name} & \colhead{z$^{a}$} & \colhead{R$_{e}$} &
\colhead{NUV$-$r} & \colhead{log$_{10}$(L$_{FIRST}$)} &  &
\colhead{Name} & \colhead{Type} & \colhead{mag$^{b}$} &
\colhead{Dist$^{c}$} & \colhead{Comment}\\
& & & \colhead{(arcsec)} & \colhead{(mag)} & \colhead{(W/Hz$^{-1}$)} &
& & & & \colhead{(kpc)} & \colhead{}
}
\startdata
1 & NGC 1260 & 0.019 & 7.76 & 5.62 & -- & & 2006gy & SNIIn & 15.0 & 0.53 & & \\
2 & SDSS J160713.55--000443.6 & 0.031 & 5.61 & 5.51 & -- & & 2001ax & SNII & 17.5 & 2.57 & & \\
3 & NGC 774 & 0.015 & 13.78 & 5.24 & -- & & 2006ee & SNII & 17.6 & 3.07 & & \\
4 & NGC 2768 & 0.005 & 29.67 & 4.63 & 20.75 & & 2000ds & SNIb & 17.9 & 2.23 & ``Ca-rich" \\
5 & SDSS J024606.79--073803.7 & 0.030 & 6.00 & 4.28 & -- & & 2008al & SNII & 17.6 & 12.89 & & \\
6 & NGC 4001 & 0.047 & 6.55 & 4.17 & -- &  & 2003ky & SNII  & 17.4 & 4.99  & & \\
7 & SDSS J153452.53+070047.9 & 0.070 & 3.21 & 3.35 & -- & & 2007ed & SNII & 19.7 & 5.73 & & \\
8 & SDSS J164734.90+495000.7 & 0.048 & 1.85 & 3.38 & 23.90 & & 2009fe & SNII & 18.1 & 0.65 & & \\
9 & SDSS J003328.04--001912.9 & 0.107 & 2.79 & 2.67 & 22.41& & 2006ho & SNII & 19.6 & 0.05 & & \\
\enddata
\tablenotetext{a}{Redshift}
\tablenotetext{b}{Apparent magnitude of the supernova}
\tablenotetext{c}{Distance between the center of the host galaxy and the supernova position}
\end{deluxetable*}
\end{center}

\section{Sample}
\label{sec:sample}

The catalogue of the Center for Astrophysics (CfA) provides all
supernovae\footnote {http://www.cfa.harvard.edu/iau/lists/Supernovae.html}
reported since 1885.  We construct a list of galaxies that have hosted SNe II
or Ibc in this catalogue that overlap with Data Release 7 of the Sloan Digital
Sky Survey (SDSS, \citealt{York00, Stoughton02, Abazajian09}).  We cross-match
the detections in the {\it GALEX} GR5/6 archive, and perform visual inspection
of both SDSS optical and {\it GALEX} UV images to select early-type galaxies.
We de-redden the colors with respect to Galactic extinction using
\citet{Schlegel98} maps provided by the SDSS pipeline and assuming
A$_{NUV}$~=~8.741~$\times$~E(B$-$V) \citep{Wyder05}. Thus, the final sample
comprises 9 early-type host galaxies of SNe II/Ib\footnote{ We exclude the SN
1986M, which is classified as SN Ib in the CfA catalogue and SN Ia in the
Asiago SN catalogue (http://web.oapd.inaf.it/supern/cat/) respectively.
Given its rather high luminosity, it is more likely to be a SN
Ia rather than a SN Ib.}. 
We also use radio observations in the Faint Images of the Radio Sky at Twenty-cm
(FIRST) survey \citep{Becker95}, which has
an angular resolution of $\sim$5'' and a completeness limit of 1 mJy. There are
3 sample galaxies with radio detections.  For comparison, we use the same
procedure to construct a sample of early-type galaxies hosting a SN Ia.

\begin{figure*}
\centering
\includegraphics[width=1\textwidth]{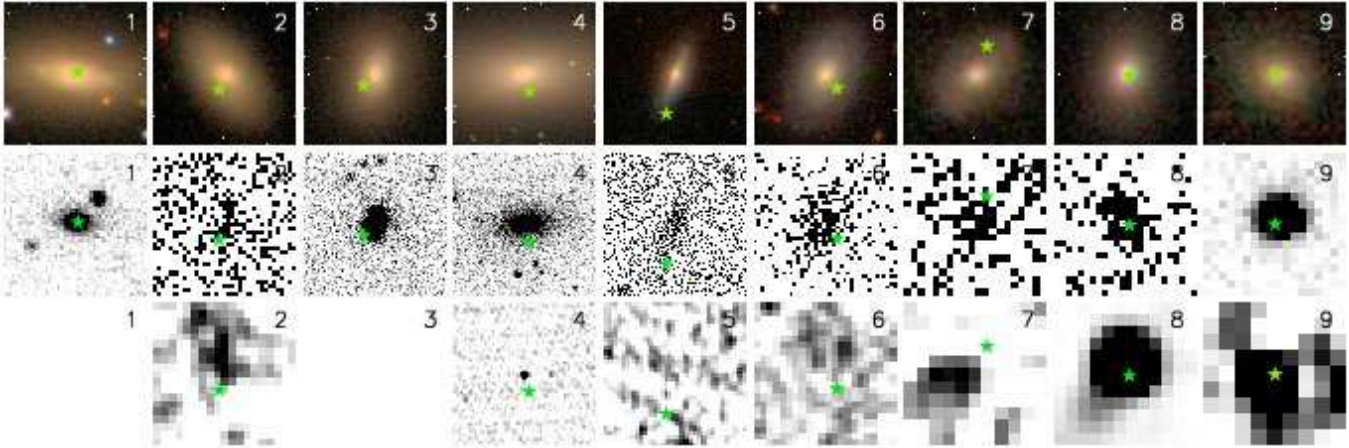}
\caption{A sample of early-type host galaxies of SN II/Ib. 
The top row is SDSS optical images, the middle row is {\it GALEX} ultraviolet images,
and the bottom row is FIRST radio images. The top optical images show the positions of
supernovae marked by star symbols. We label the host galaxy listed in Table~\ref{tbl:sample}
in each panel.}
\label{fig:img}
\end{figure*}

We list the SNe II/Ib sample in Table~\ref{tbl:sample}, which includes one
confirmed SN Ib and eight SNe II. All SNe II in our sample (except for SN 2006gy)
seem to have luminosity typical of core collapse supernovae (see
Table~\ref{tbl:sample}); thus, they are not likely to belong to a hybrid
class of SN Ia. Figure~\ref{fig:img} shows a sample of early-type host
galaxies of SNe II/Ib with optical, UV, and radio images.  The SDSS optical
images are shown in the top row, where the positions of SNe are marked by star
symbols.  The {\it GALEX} UV images and the FIRST radio images are shown   in
the second and third rows, respectively.  Most of the SN II/Ib reside close to the
center of the host galaxy, except for SN 2008al. Their locations are similar to those
indicated by UV and radio emission.

Regarding morphology of host galaxies, NGC 1260 (label 1) and NGC 774 (label 3)
are classified as S0 in the NASA/IPAC Extragalactic Database (NED).  NGC 2768
(label 4) is classified as elliptical.  Labels 5, 7, 8, and 9 have {\texttt
fracDev}=1, which is the weight of de Vaucouleurs profile in the best composite
(de Vaucouleurs + exponential) fit to the image in r band.  Labels 2 and 6 have
{\texttt fracDev}=0.89.  All nine galaxies have concentration indices greater than
2.5.  All galaxies are fairly bright, and all but one are close ($z < 0.05$); thus,
morphology determination based on visual inspection and pipeline
parameters is deemed reliable.

\section{UV$-$optical color-magnitude relation}
\label{sec:uvcmr}

\begin{figure}
\centering
\includegraphics[width=0.5\textwidth]{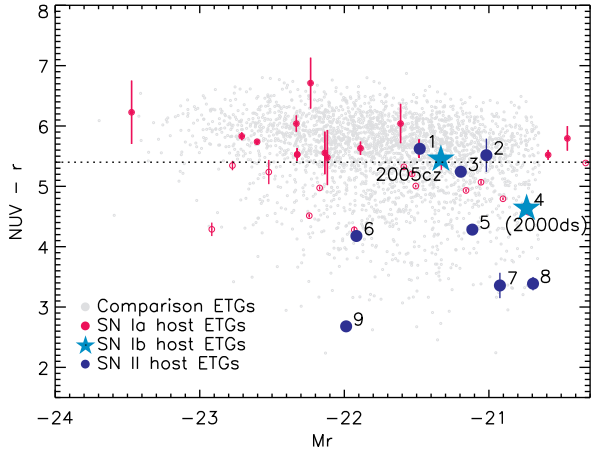}
\caption{The UV$-$optical color-magnitude relation. We plot early-type host galaxies of
SNe II and SNe Ia with blue and red circles, respectively. 
The dotted line shows the NUV$-$r = 5.4 cut, which indicates the color of the nearby
UV-upturn galaxy, NGC~4552 (see the text for details).
We mark the SN Ia hosts with a color above/below this cut by filled/open red circles.
Star symbols denote SN Ib host galaxies.
The grey points are the comparison sample of early-type galaxies from SDSS-GALEX cross-matches
(that is, without a detected SN).
We also label each host galaxy listed in Table~\ref{tbl:sample}.}
\label{fig:cmr}
\end{figure}

The UV$-$optical color magnitude relation (see, e.g., \citealt{Yi05,
Kaviraj07,S07a}) is a particularly efficient tool for studying recent star
formation in early-type galaxies because of its high sensitivity to young
stellar populations. Figure~\ref{fig:cmr} shows the NUV$-$r color
magnitude relation for the SN II host galaxies (blue circles with label), the
SN Ib (2000ds) host galaxy (star symbol), and the SN Ia host galaxies (small red
circles).  We also show the host galaxy of SN 2005cz (NGC 4589; cf.
\citealt{Kawabata10}), with a star symbol.  We do not have SDSS photometry of
this galaxy, so we convert its Johnson magnitudes (from NED) to SDSS
magnitudes using the equations given by \citet{Jester05}.  We also show typical
early-type galaxies (grey dots) for comparison.  The empirical threshold of
NUV$-$r $<$ 5.4 (dashed line) is a criterion that is used to find old galaxies
with recent star formation.  \citet{Yi05} derived this value from the nearby
prominent UV-upturn galaxy NGC 4552; hence, it should be considered a
lower bound in the NUV$-$r color of purely-old stellar populations. 

A cursory inspection of this diagram shows that most of the early-type host
galaxies of SNe II/Ib reside below NUV$-$r $\sim$ 5.4, implying that they have
undergone  recent star formation, while the SN Ia host galaxies show relatively
redder UV$-$optical colors. We also find that SNe II/Ib preferentially occur in
relatively small host galaxies ($M_\mathrm{r} \ga -22$) compared to the case of
SNe Ia \citep[cf.][]{Arcavi10}.
This is not surprising considering that supernova rates correlate with star formation rates.
It should be noted that less massive
early-type galaxies are more likely to have experienced recent star formation \citep{S07a,Jeong09}.
\citet{Li10} finds that SNe II preferentially occur in relatively smaller host galaxies.
But direct comparison is difficult because interpretation on galaxy size can be tricky
when dealing with a heterogeneous sample of galaxy morphologies.

The host galaxy of SN 2000ds (NGC 2768, label 4) shows a NUV$-$r color that is
significantly bluer than those of most SN Ia hosts, strongly implying the
presence of young stars.  On the other hand, NGC 4589 (star symbol), which is the host
galaxy of another peculiar SN Ib (SN 2005cz), shows a marginal NUV$-$r color.
Note that NGC 4589 is a strong 21 cm HI source
\citep{Theureau07} and shows large opaque dust patches \citep{MM94, Carollo97,
Lauer05}. Also note that NUV colors are more sensitive to dust than optical colors by a factor of 4,
so the presence of a small amount of dust can have a proportionally
greater effect in the NUV than in the optical colors. After all, the intrinsic
colors of this galaxy might be substantially bluer than the observed colors.

Our sample includes the host galaxy of SN 2006gy (NGC 1260, label 1), which is
one of the most luminous supernovae ever discovered.  Even though the
progenitor of this highly luminous SN IIn is generally believed
to be a very massive star \citep[e.g.,][]{Smith10}, the NUV$-$r color of the
host galaxy is not particularly blue compared to the other SN II/Ib host
galaxies. Like NGC 4589, this galaxy has also been found to be fairly dusty
\citep{Ofek07}. 

We find that the host galaxy of SN 2000ds and two of the eight SN II host galaxies
that show a markedly bluer NUV$-$r color (labels  8 and 9) are strong radio
sources (see Table~\ref{tbl:sample} and Figure~\ref{fig:img}). Recently, some
authors argued that radio loudness in early-type galaxies may be related to a
recent merger event that enhanced the SN Ia rate from young stellar
populations~\citep{DV05, Mannucci06, Graham10}.  If this is the case,  the
strong radio emission in these galaxies would serve as more evidence of
recent star formation in these host galaxies. However, not all SN II
host galaxies in our sample have significant radio emission. The issue of
the correlation between radio emission and the presence of young stellar 
populations should be addressed carefully in future work. \\


\section{Stellar population analysis}

The use of NUV$-$r color to find galaxies with recent star formation (as
discussed in the previous section) is overly simplistic, as it is
based on the assumption that all early-type galaxies should have
an underlying UV-upturn component from old stars at a level similar to that
of NGC 4552 \citep[see the review
of][]{OConnell99}. Contrarily, it was recently found that only a small
fraction of early-type galaxies show a UV upturn \citep{Yi10}.  Hence,
we hereby use a more sophisticated method to characterize the stellar
populations in our SN host galaxies.

We consider a two-stage star formation history to determine the age and mass
fraction of the young stellar component in the SN host galaxies.  Both
starbursts are assumed to be instantaneous.  The base models in this study are
taken from \citet{Yi03}, which are specialized for old stellar populations.
Since these models do not cover ages younger than 1~Gyr,  we combine
them with the models of \citet{BC03} for a young stellar population
\citep{fs00, Kaviraj07, S07b}.  We assume a single uniform age of 12~Gyr for
the old component.  The young component is allowed to vary in age ($10^{-3} \le
t_{young} \le $10~Gyr) and mass fraction ($10^{-6} \le f_{young} \le 1$).  More
complex models, such as those based on multiple star bursts or continuous star
formation, do not affect this analysis very much because UV lights are
insensitive to the details of the star formation history, but are highly dominated
by the most recent star formation event.

\begin{figure}
\centering
\includegraphics[width=0.5\textwidth]{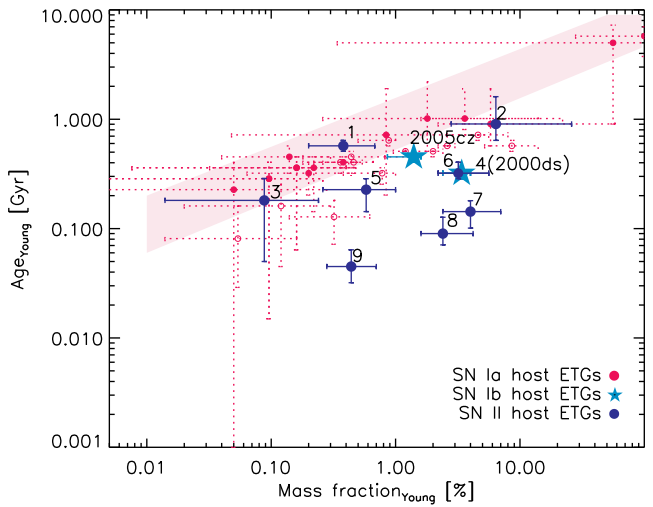}
\caption{The distribution in age and mass fraction for the young stellar component.
Blue and red symbols represent the best fit of early-type host galaxies of
SNe II and SNe Ia, respectively. Star symbols indicate the SN Ib host galaxies.
The pink-shaded region marks the age-mass degeneracy area. See the text for details. 
}
\label{fig:tyfy}
\end{figure}

To constrain these two parameters, we fit the observed UV and optical colors to
models, and compute the associated $\chi^{2}$ statistic to obtain a probability
distribution of the age and mass fraction of the young stellar component.
Internal extinction is also constrained in the $\chi^{2}$
minimization.  The best fit and confidence levels are shown in
Figure~\ref{fig:tyfy}.  The $\chi^{2}$ minimum is marked with circles along
with error bars. 

We first note that this analysis is heavily subject to various degeneracies.
The most important one for the UV analysis is the age-mass degeneracy.  A
larger fraction of the young component is hardly distinguishable from a smaller
fraction of the younger component. 

The majority of SN Ia host galaxies have a characteristic age of $3 -
6~10^{8}~\mathrm{yr}$ for the young stellar component, which agrees with
previous studies \citep[e.g.][]{Mannucci05, Sullivan06, S09}.  Our
{\it quiescent} SN Ia host galaxies (filled red circles) are found to have a
small amount of young stars, but this is because our building-block population
models do not have a significant UV flux such as that exhibited by UV upturn
galaxies (e.g., NGC~4552). These galaxies usually have a small UV flux,
with which robust estimations of the young component properties are difficult.
They lie along the degeneracy area marked by the pink-shaded region.  In this
regard, the young component mass fraction is an upper limit in the case of SN
Ia host galaxies.  Note that the SN Ia host galaxies with a blue UV-optical
color (open red circles) are mostly below this shade, indicating either a
greater amount of young stars or a younger age, both of which produce a higher
UV flux.

The SN II/Ib host galaxies are markedly different from the {\it quiescent} SN Ia
host galaxies.  Most of the SN II/Ib host galaxies are located below the
degeneracy region, strongly indicating the presence of young stars. \\

\section{Implications for the progenitors \\ of faint, Ca-rich SNe Ib}

The SNe II in our sample must have a core-collapse origin as mentioned in
\S~\ref{sec:sample},  and the above discussion confirms that their progenitor
ages are systematically smaller than those of the SNe Ia.  The host galaxy of SN
2000ds (NGC 2768)  appears to have properties similar to those of the SNe II
host galaxies.  The relatively strong radio emission of NGC 2768 may be
additional evidence to support  the connection between SN 2000ds and recent
star formation.  The host galaxy of SN 2005cz (NGC4589) also seems to show
signatures of recent star formation.  It is  located below the degeneracy
region in Figure~\ref{fig:tyfy}, and known to be dusty as discussed above.  Our
result thus indicates that the faint, Ca-rich SNe Ib 2000ds and 2005cz
originate from massive stars like the SNe II in our sample. 

Our result may give some hints for their progenitor masses.  Given the characteristic age
of about 300 - 400 Myr for the young stellar component in NGC 2768 and NGC 4589
according to our stellar population model, the progenitors of SN 2000ds and SN
2005cz are not likely to have an initial mass much larger than about
$10~M_\odot$ This conclusion is very similar to that of \citet{Kawabata10} for
SN 2005cz. 
They argued that the high Ca to Oxygen
abundance ratio implied by observation in the ejecta of such peculiar SNe Ib can be best explained by
explosive nucleosynthesis in a relatively small stellar core containing very thin silicon and oxygen shells.
The corresponding helium core size may not significantly exceed $3~M_\odot$.
This independently points to moderately massive progenitors having initial masses of about $10 M_\odot$. 

Both being SNe Ib, the progenitors of SN 2000ds and SN 2005cz must have
lost their hydrogen envelope, leaving naked helium cores of $2-3~M_\odot$.
This could be only possible in close binary systems,
unless the mass loss rate of their progenitors were unrealistically high
\citep[e.g.,][]{Kawabata10, Yoon10}.  It is important to
note that the initial masses of core collapse supernovae could be lower down
to about $5-6~M_\odot$ with binary interactions, compared to the canonical
limit of $8-10~M_\odot$ for single stars.  Such binary star evolution
with initial masses of $5 - 11 M_\odot$ towards  helium stars of about
$2-3~M_\odot$ such as SN Ibc progenitors is \emph{required} to explain the formation of
binary pulsars PSR B2303+46 and PSR J1141-6545 with old white dwarf
companions \citep[e.g.][]{Tauris00}.  Therefore, it is not surprising
that SNe Ib like 2000ds can be found in some early-type galaxies, hovering around
relatively young stars with ages of $\sim 10^8$ Myr. \\


\section*{Acknowledgments}
This project made use of the SDSS optical data, the {\it GALEX} ultraviolet data, the HyperLeda database
and the NASA/IPAC Extragalactic Database.
This work was supported by the National Research Foundation of Korea 
through the Doyak grant (No. 20090078756) and the SRC grant to the Center for Galaxy Evolution Research.

\end{document}